\documentstyle[12pt]{article}
\begin{document}

\centerline {\Large Stick-slip statistics for two fractal surfaces:}
\centerline {\Large A model for
earthquakes}

\bigskip

\centerline {Bikas K. Chakrabarti$^{*,+}$ and Robin B. Stinchcombe$^*$}

\centerline {$^*$Theoretical Physics, Department of Physics, University of
Oxford}
\centerline {1 Keble Road, Oxford, OX1 3NP, England.}

\centerline {$^+$Saha Institute of Nuclear Physics}
\centerline {1/AF Bidhannagar, Calcutta 700064, India.}

\bigskip

\bigskip

\noindent {\bf Abstract:}  
Following the observations
of the self-similarity in various length scales in the roughness of
 the fractured solid surfaces, we propose here a new model for the
earthquake. We demonstrate rigorously that the contact area
distribution between two fractal surfaces follows an unique power
law. This is then utilised to show that the elastic energy releases
for slips between two rough fractal surfaces indeed follow a 
Guttenberg-Richter like power law.

\bigskip

\noindent {\bf 1. Introduction}

\medskip

The earth's solid outer crust, of about 20 kilometers in 
average thickness, 
rests on the tectonic shells. Due to the high temperature-pressure
phase changes and the consequent powerful convective flow in
the earth's mantle, at several hundreds of kilometers of depth,
the tectonic shell, divided into a small number (about
ten) of  mobile plates, has relative velocities of the order of a 
few centemeters per year [1,2]. Over several tens of years, 
enormous elastic strains develop sometimes on the earth's
crust when sticking (due to the solid-solid friction) to the
moving tectonic plate. When slips occur between the crust and
the tectonic plate, these stored elastic energies are 
released in 'bursts', causing the damages during the earthquakes.
Because of the uniform motion of the tectonic plates, the 
elastic strain energy stored in a portion of the crust (block),
moving with the plate relative to a 'stationary' neighbouring portion
of the crust,  can vary only due to the random strength of
the solid-solid friction between the crust and the plate. The slip
occurs when the accumulated stress exceeds the frictional force.
As mentioned  before, the observed distribution of the elastic
energy release in various earthquakes seems to follow a power
law. The number of earthquakes $N(m)$, having magnitude in the 
Richter scale greater than or equal to $m$, is phenomenologically
observed to decrease with $m$ exponentially.
This gives the Guttenberg-Richter law [1]

$${\rm ln} N(m) = {\rm constant} - a~ m, \eqno (1)$$

\noindent where $a$ is a constant. It appears [1,2] that the amount of 
energy $\epsilon$ released in an earthquake of magnitude $m$
is related to it exponentially:

$$ {\rm ln} \epsilon = {\rm constant} + b~m, \eqno (2)$$

\noindent where $b$ is another constant. Combining therefore we get the
power law giving the number of earthquakes $N(\epsilon)$
releasing energy greater than or equal to $\epsilon$ as

$$ N(\epsilon) \sim \epsilon^{-\alpha}, \eqno (3)$$

\noindent with $\alpha = a/b$. For most regions of the earth $a \simeq
1$. The estimates of $b$ are more difficult, and vary 
between 1.0 and 1.5 [1,2]. Hence, the observed value of the
exponent (power) $\alpha$ in (3) ranges between 0.7 and 1.0.
 
 Several laboratory and computer simulation models have 
recently been proposed [3]  to capture essentially the above
power law in the earthquake energy release statistics. In a
table-top laboratory simulation model of earthquakes, Burridge
and Knopoff [4] took a chain of wooden blocks connected by identical
springs to the neighbouring blocks. The entire chain was
placed on a rigid horizontal table with a rough surface, and
one of the end blocks was pulled very slowly
and uniformly using a driving 
motor. The strains of the springs increase due to the creep motions 
of the blocks until one or a few of the blocks slip. The drops
in the elastic energy of the chain during slips could be 
measured from the extensions or compressions of all the springs, and
could be taken as the released energies in the earthquake. For some 
typical roughness of the surfaces (of the blocks and of the table),
the distribution of these drops in the elastic energy due to slips
indeed shows a power law behaviour with $\alpha \simeq$ 1 in (3). 
A computer simulation version of this model by Carlson and
Langer [5]  considers harmonic springs connecting equal mass
blocks which are also individually connected to a rigid frame
(to simulate other neighbouring portions of the earth's crust
not on the same tectonic plate) by harmonic springs. The entire
system moves on a rough surface with nonlinear velocity dependent 
force (decreasing to zero for large relative velocities) in the
direction opposite to  the relative motion between the block and
the surface. In the computer simulation of this model it is seen
that the distribution of the elastic energy release in such
a system can indeed be given by a power law like (3), provided the
nonlinearity of the friction force, responsible for the 
self-organisation, is carefully chosen [5]. The lattice automata model
of Bak et al [6]  and its modification appropriate to
the Burridge-Knopoff model by Olami et al [6] 
represent the stress on each block by a height variable
at each lattice site. The site topples (the block
slips) if the height (or stress) at
that site exceeds a preassigned threshold value, and the height becomes 
zero  there and the neighbours share the stress by increasing their
heights by one unit. With this dynamics for the system,  if
any of the neighbouring sites  of the
toppled one was already at the threshold height,
the avalanche continues. The boundary sites are considered to be all
absorbing. With random addition of heights at a constant rate
(increasing stress at a constant rate due to tectonic motion), such
a system reaches its self-organised critical point where the
avalanche size distributions follow a natural power law corresponding
to this self-tuned critical state. Bak et al [6]  identify 
this self-organised critical state to be responsible for the
Guttenberg-Richter type power law. All these models are successful in
capturing the Guttenberg-Richter power law, and the real reason
for the self-similarity inducing the power law is essentially the
same in all these different models: emergence of the self-oganised 
critical  state for wide yet suitably chosen variety of nonlinear
many-body coupled dynamics. In this sense all these models incoporate
the well-established fact of the stick-slip frictional instabilities
between the earth's crust and the tectonic plate. It is quite 
difficult to check at this stage any further details and predictions
of these models.

While the motion of the tectonic plate is surely an observed fact, 
and this stick-slip process should be a major ingredient of any
bonafide model of earthquake, another established fact regarding
the fault geometries of the earth's crust is the fractal nature 
of the roughness of the surfaces of the earth's crust and the
tectonic plate. This latter feature is
 missing in any of these models discussed
above.  In fact, the surfaces involved in the process are
results of large scale fracture seperating the crust from
the moving tectonic plate. Any such crack surface is observed
to be a self-similar fractal, having the self-affine scaling
property $z(\lambda x, \lambda y) \sim \lambda^{\zeta} z(x,y)$
for the surface coordinate $z$ in the direction perpendicular to 
the crack surface in the ($x,y$) plane [7]. Various fractographic
investigations indicate a fairly robust universal behaviour
for such surfaces and the roughness exponent $\zeta$ is observed to have
a value around 0.80-0.85 (with a possible crossover to $\zeta
\simeq$ 0.4 for slow propagation of the crack-tip) [8,3]. This widely
observed scaling property of the fracture surfaces also suggests that
the fault surfaces of the earth's crust or the tectonic plate 
should have similar fractal properties. In fact, some investigators 
of the earthquake dynamics have already pointed out that the fracture
mechanics of the stressed crust of the earth forms self-similar
fault patterns, with well-defined fractal dimensionalities near
the contact areas with the major plates [9]. Although no realistic
models have been developed incorporating this
 fact, these investigators [9]
have often pointed out that such self-similarities of the 
fault surfaces could be responsible  for the Guttenberg-Richter
power law (3). 

\bigskip

\noindent {\bf 2. Model and RG calculations}

\medskip

In our model, the solid-solid  contact surfaces of both the earth's
crust and the tectonic plate are considered as average self-similar
fractal surfaces. We then consider the distribution of contact
areas, as one fractal surface slides over the other.
We relate the total contact area between the two surfaces to
be proportional to the elastic strain energy that can be grown
during the sticking period, as the solid-solid friction force
arises from the elastic strains at the contacts
 between the asperities [10].
 We then consider this energy to be released as one surface
slips over the other and sticks again to the next contact or
overlap between the rough fractal surfaces.
Considering that such slips occur at intervals  proportional
to the length corresponding to that area, we obtain a power
law for the frequency distribution of the energy releases. This 
compares quite well with the Guttenberg-Richter law.

In order to proceed with the estimate of the number density 
$n(\epsilon)$ of earthquakes releasing energy $\epsilon$ 
in our model, we first find out the distribution  
$\rho (s)$ of the overlap or contact area $s$ between two 
self-similar fractal surfaces. We then relate $s$ with $\epsilon$
and the frequency of slips as a function of $s$, giving finally
 $n(\epsilon)$. To start with a simple problem of contact
area distribution between two fractals, we first take two Cantor
sets [11]  to 
model the contact area variations of two (nonrandom and self-similar)
surfaces as one surface  slides
over the other. Figure 1(a) depicts structure in such surfaces
at a scale which corresponds to only the second generation of
iterative construction of two displaced Cantor sets, shown 
in Fig. 1(b). It is obvious that with successive iterations, 
these surfaces will acquire self-similarity at every length
scale, when the generation number goes to infinity. We intend to
study the distribution of the total overlap $s$ (shown by the
shaded regions in Fig. 1(b)) between the two Cantor sets, in
the infinite generation limit. Let the sequence of generators 
$G_l$ define our Cantor sets within the interval [0,1]: $G_0
= [0,1], G_1  \equiv RG_0 = [0,a] \bigcup
[b,1] $ (i.e., the union of the intervals $[0,a]$ and $[b,1]),
~ ...~, G_{l+1} = RG_l, ~ ... ~$. If we represent the
mass density of the set $G_l$ by $D_l(r)$, then $D_l(r)$ = 1 if $r$
is in any of the occupied intervals of $G_l$, and $D_l(r)$ = 0 
elsewhere. The required overlap magnitude between the sets
at any generation $l$ is then given by the 
convolution form  $s_l(r) = \int dr'
D_l(r') D_l(r-r')$. This form applies to symmetric fractals (with
$D_l(r) = D_l(-r)$); in general the argument of the second $D_l$
should be $D_l(r+r')$. 

Some aspects of the convolution of two Cantor sets
has previously been discussed [12] in connection with
band-width and band number transitions in quasicrystal
models. The generalisation given hereunder of the
earlier employed recursive scaling method provides a 
very direct solution to the more complex problem
we encounter here. One can
express the overlap integral $s_1$ in the first generation by 
the projection  of the shaded regions along the 
vertical diagonal in Fig. 2(a). That gives the form 
shown in Fig. 2(b).
 For $a=b \le {1\over 3} $, the
nonvanishing $s_1(r)$ regions do not overlap, and 
are symmetric on both sides with the slope of the middle
curve being exactly double  those on the sides. One can then
easily check that the distribution $\rho_1(s) $ of  overlap $s$ at this
generation is given by Fig. 2(c), with both $c$ and $d$ greater
than unity, maintaining the normalisation of the probability
$\rho_1$ with $cd = 5/3$. The successive generations of the 
density $\rho_l(s)$ may thefore be represented by Fig. 3,
where

$$ \rho_{l+1}(s) = \tilde R \rho_l(s) \equiv {d\over 5}
\rho_l \left({s\over c}\right) + {4d\over 5}\rho_l \left(
{2s\over c}\right). \eqno (4)$$

\noindent In the infinite generation limit of the
renormalisation group (RG) equation, if $\rho^*(s) $ denotes
the fixed point distribution such that $\rho^*(s) = \tilde R
\rho^*(s)$, then assuming $\rho^*(s) \sim s^{-\gamma} \tilde
\rho(s)$, one gets $(d/5)c^{\gamma} + (4d/5)(c/2)^{\gamma} =$
1. Here $\tilde \rho(s)$ represents an arbitrary modular
function, which also
includes a logarithmic correction for large $s$. This
agrees with the above mentioned normalisation condition 
$cd = 5/3$ for the choice $\gamma = 1$.   This result 
for the overlap distribution

$$ \rho^*(s) \equiv \rho(s) \sim s^{-\gamma}; ~~ \gamma =1,
\eqno (5)$$

\noindent  is the general result [13] for all cases that
we have investigated and solved by the functional rescaling
technique (with the $\log s$ correction for large $s$,
renormalising the total integrated distribution).
 They include overlaps of different non-random 
Cantor sets, as well as higher dimensional fractals (for
slides along various directions) like the carpet whose
first generation is shown in Fig. 4. Also, by
combining weighted probabilities for two such nonrandom fractals,
one can generate recursion relations like (4) for the overlap
distribution function for random fractals, which
is the realistic case. One again gets the
same result (5), except that the 
 modular part $\tilde \rho(s)$ is periodic or quasi-periodic
for general non-random fractals, and is normally constant
for the random fractals. For all these cases, one 
gets the general result (5) for all the fractal surfaces. An
obvious exception is where either (or both) of the surfaces is
compact (Euclidean); there 
the distribution retains the original form,  
Fig. 3(a), in the successive generations
and consequently gives $\gamma$ = 0. It may be mentioned here 
that, in the context of flow through fractured media (underground
mineral oil recovery), considerable investigations have already
been made [14]  on  the scaling behaviour in the aperture
of the channels between two rough fractured solid surfaces, even as
one surface slides over the other. We investigate here a different
aspect of the problem, namely the contact area distribution of
such  such surfaces, as one slides over the other.

It may now be argued that the elastic energy
$\epsilon$  stored between the two microscopically rough
surfaces will be approximately proportional to the contact area
$s$ between the two surfaces. This is because during the sticking
period, the solid-solid friction arises due to the elastic 
forces caused by elastic strains of the contact asperities of the
surfaces [10].  Of course, the densities of slips of size $s$
are not evenly distributed. As the tectonic plate moves with a constant
velocity, we can assume that for a slip of contact area $s$ the
typical distance to be moved, and hence the time taken, will be
of the order of $s^{\delta}$, with $\delta = 1/D$ where $D$ is
the fractal dimensionality of the surfaces. This suggests that
for a finite-time-average statistics, the number density $n(\epsilon)$
of quakes releasing energy $\epsilon$ will be given by

$$n(\epsilon) \sim \epsilon^{-(\gamma + \delta)}, \eqno (6)$$

\noindent where $\gamma = 1$, and $\delta = 1/D$. This in turn implies
that the number density $N(\epsilon)$ of quakes releasing energy
greater than $\epsilon$ is given by

$$N(\epsilon) = \int_{\epsilon}^{\infty} n(\epsilon') d\epsilon' \sim
\epsilon ^{-\alpha}; ~~~ \alpha = \gamma +\delta -1 = \delta,
\eqno (7) $$
with $\delta = 1/D$. This gives us the Guttenberg-Richter power law
(3) with $\alpha = 1/D$. As the
contact-surface fractal dimension $D$ (given
by the roughness exponent $\zeta$ discussed earlier) is generally
less than two, our theory predicts the value of $\alpha$ to be
greater than 0.5. This may be compared with its observed value
ranging between 0.7 and 1.0.

\bigskip

\noindent {\bf 3. Discussions}

\medskip

As emphasized already, we agree with the physicists' identification
of the Guttenberg-Richter law (3) as an extremely significant 
one in geophysics. Like the previous attempts [4-6], we also
develop here a model to capture this important feature in its
resulting statistics. Judging from the comparisons
of the exponent values $\alpha$ in (3) and (7),
the model succeeds  at least as well as
 the earlier ones. More
importantly, our  model incorporates 
both the geologically observed facts: fractal nature of the 
contact surfaces of the crust and of the tectonic plate [7-9],
 and the
stick-slip motion between them [1,2]. However, the origin of the power
law in the quake statistics here is the self-similarity of the
fractal surfaces, and not any self-organisation directly in their 
dynamics. In fact, the extreme non-linearity in the nature of the
crack propagation [3,8] is responsible for the fractal nature
of the rough crack surfaces of the crust and the tectonic plate. This
in turn leads  here to the Guttenberg-Richter like power law  in
the earthquake statistics.

\bigskip

\noindent {\bf Acknowledgements:} BKC is grateful to the Exchange
Programme between the Indian National Science Academy and the 
Royal Society for supporting his visit to the Physics Department,
Oxford University, Oxford, where this work was done. We are grateful
to G. Ananthakrishna, A. Hansen, H. J. Herrmann, D. Fisher,
S. S. Manna  S. Roux  for J. P. Vilotte for useful
information, comments and criticisms.

\bigskip

\bigskip

\noindent {\bf References:}

\medskip

\noindent [1] B. Guttenberg and C. F. Richter,    {\it Seismicity 
of the Earth and Associated Phenomena}, Princeton Univ. Press,
Princeton, N.J. (1954).

\medskip

\noindent [2] B. V. Kostrov and S. Das,  {\it Principles of 
Earthquake Source Mechanics}, Cambridge Univ. Press, Cambridge (1988);
C. H. Scholz,  {\it The Mechanics of Earthquake and Faulting},
Cambridge Univ. Press, Cambridge (1990).

\medskip

\noindent [3]  B. K. Chakrabarti and L. G. Benguigui,  {\it 
Statistical Physics of Fracture and Breakdown in Disordered
Solids}, Oxford Univ. Press, Oxford (1997). 

\medskip

\noindent [4] R. Burridge and L. Knopoff,  {\it Bull.
Seis. Soc. Am.} {\bf 57} 341-371 (1967).

\medskip

\noindent [5] J. M. Carlson and J. S. Langer, {\it Phys. Rev. Lett.}
{\bf 62} 2632-2635 (1989); J. M. Carlson, J. S. Langer and B. E. Shaw, 
{\it Rev. Mod. Phys.} {\bf 66} 657-670 (1994).

\medskip

\noindent [6] P. Bak, C. Tang and K. Weisenfeld, {\it  Phys. Rev.
Lett.} {\bf 59} 381-384 (1987); P. Bak and C. Tang, {\it J. Geophys. Res.}
{\bf 94} 15635-15637; Z. Olami, H. J. S. Feder and K. Christensen, 
{\it Phys. Rev. Lett.} {\bf 68} 1244-1247 (1992).

\medskip

\noindent [7] B. B.  Mandelbrot, D. E. Passoja and A. J.  Pullay, 
{\it Nature} {\bf 308} 721-722 (1984).

\medskip

\noindent [8] J. P. Bouchaud,  E. Bouchaud,  G. Lapasset and J. Planes,
{\it Phys. Rev. Lett.} {\bf 71} 2240-2243 (1993); Marder, M. \& Feinberg, J. 
{\it Phys. Today}, September, 24-29 (1996); Chakrabarti and B. K. Chakrabarti,
{\it Physica A} (this Proc. Vol., 1999).

\medskip

\noindent [9] Y. Y. Kagan,  {\it Geophys. J. Royal Astrophys. Soc.
Canada} {\bf 71} 659 (1982); B.  Barriere,  and D. L. Turcotte,  
{\it Geophys. Res. Lett. } {\bf 18} 2011-2014 (1991); M. Sahimi,
 M. Robertson, {\it Physica A} {\bf 191} 57-68 (1992).

\medskip

\noindent [10] C. Caroli and Ph. Nozieres,  {\it Eur. Phys. J. B}
{\bf 4} 233-246 (1998).

\medskip

\noindent [11] B. B. Mandelbrot,  {\it Fractal Geometry of Nature},
Freeman Press, New York (1982).

\medskip

\noindent [12] J. A. Ashraff, J. -M. Luck and R. B. Stinchcombe, {\it 
Phys. Rev. B} {\bf 41} 4314-4329 (1990) -- Appendix B considers the 
Cantor Set convolution.

\medskip

\noindent [13] Such results can be checked by dealing explicitly
with the large $l$ behaviour of an apprropriately rescaled version
of the actual $\rho_l(s)$ resulting from $\tilde R^l \rho_0(s)$.

\medskip

\noindent [14] F. Plourabou${\acute {\rm e}}$, P. Kurowski, 
J. P. Hulin,  S.
Roux, and J. Schmittbuhl, {\it Phys. Rev. E} {\bf 51} 1675-1685
(1995), and references therein.

\bigskip

\bigskip

\noindent {\bf Figure Captions:}

\medskip

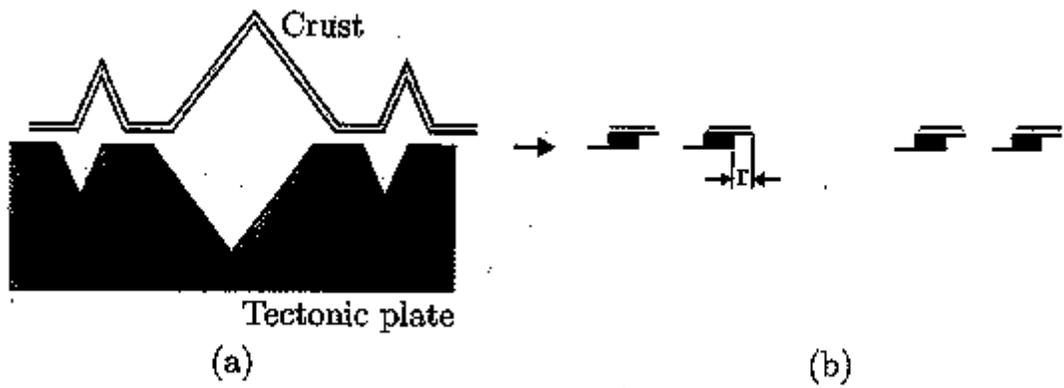
\noindent Fig. 1. (a) Schematic representations of a portion of the
rough surfaces of the earth's crust and the supporting (moving)
tectonic plate. (b) The one dimensional projection of the surfaces
form Cantor sets of varying contacts or overlaps as one surface
slides over the other.

\medskip

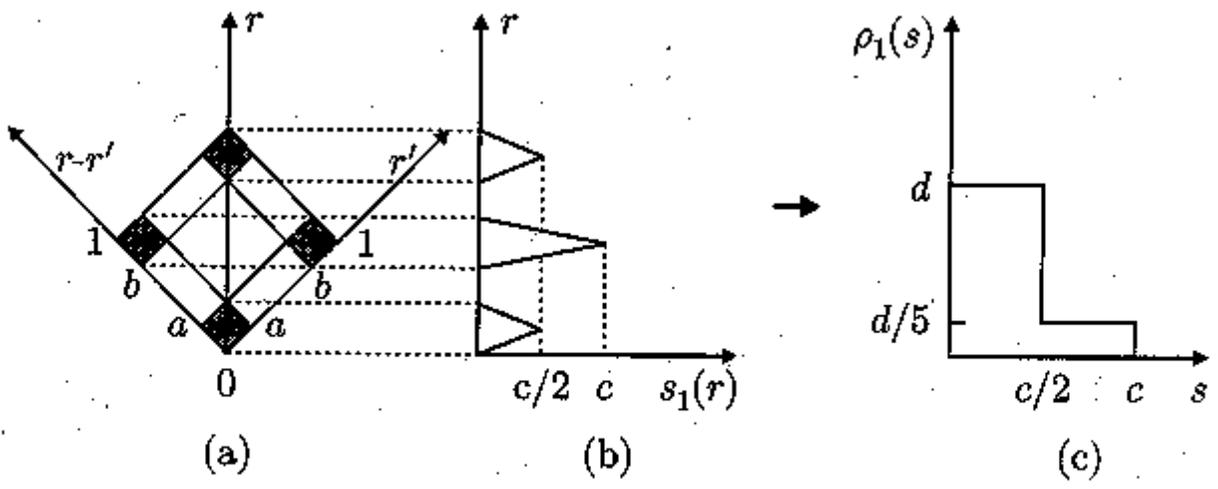
\noindent Fig. 2. (a) Two cantor sets (in their
first generation) along the axes $r$ and $r-r'$. (b) This gives
the overlap $s_1(r)$ along the diagonal. (c) The corresponding
density $\rho_1(s)$ of the overlap $s$ at this generation.

\medskip

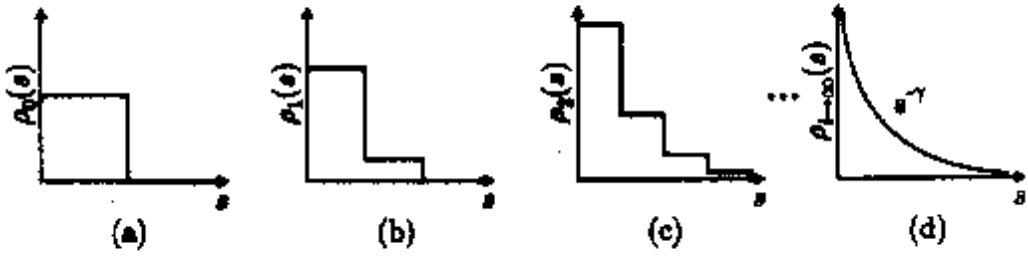
\noindent Fig. 3. The overlap densities $\rho (s)$ at various
generations of the Cantor sets: at the zeroth (a), first (b),
second (c) and at the infinite (or fixed point) (d) generations.

\medskip

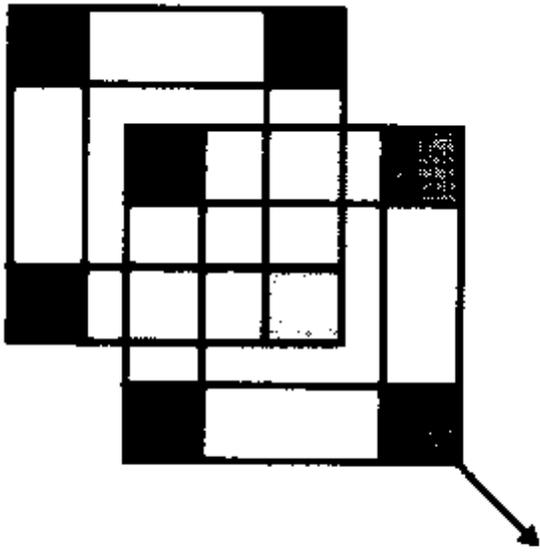
\noindent Fig. 4. Slides of two fractal crapets. The shaded
regions indicate the material content portions of the surfaces
at the second generation of an iterative construction of a fractal
surface which is a Cantor set in each of the orthogonal
directions.
The overlap of the two carpets, for an  arbitrary slide of one  
over the other, can be calculated by generalisation of the
technique used to obtain (4).

\end{document}